\documentclass[achemso,preprint,a4paper,superscriptaddress]{achemso}
\usepackage[T1]{fontenc}
\usepackage[utf8]{inputenc}

\usepackage{amsmath}
\usepackage{amsfonts}
\usepackage{amssymb}
\usepackage{mhchem,siunitx}
\usepackage{multirow}
\usepackage{caption}
\usepackage{changes}
\newcommand*{\citen}[1]{%
  \begingroup
    \romannumeral-`\x 
    \setcitestyle{numbers}%
    \cite{#1}%
  \endgroup   
}

\usepackage{graphicx}
\usepackage{color}
\usepackage{mathtools}
\usepackage{hyperref}
\usepackage{epstopdf}
\usepackage[bottom]{footmisc}
\usepackage{natbib}
\hypersetup{ 
	colorlinks=true,
	urlcolor=black,citecolor=black, linkcolor=black}

\title{Effects of mixing and annealing on the optical and mechanical properties of \ce{TiO2}:\ce{Ta2O5} amorphous coatings}

\author{Alex Amato}
\affiliation{Université de Lyon, Université Claude Bernard Lyon 1, CNRS, Institut Lumière Matière, F-69622 Villeurbanne, France} 
\alsoaffiliation{Current affiliation: Maastricht University, P.O. Box 616, 6200 MD Maastricht, The Netherlands} 
\alsoaffiliation{Current affiliation: Nikhef, Science Park 105, 1098 XG Amsterdam, The Netherlands}
\altaffiliation{These authors contributed equally to this work.} 
\email{a.amato@maastrichtuniversity.nl}

\author{Michele Magnozzi}
\affiliation{OptMatLab, Dipartimento di Fisica, Universit\`{a} di Genova, via Dodecaneso 33, 16146 Genova, Italy} 
\alsoaffiliation{Istituto Nazionale di Fisica Nucleare, Sezione di Genova, via Dodecaneso 33, 16146 Genova, Italy} 
\altaffiliation{These authors contributed equally to this work.} 
\email{magnozzi@fisica.unige.it}

\author{Nikita Shcheblanov}
\affiliation{Navier, Ecole des Ponts, Univ Gustave Eiffel, CNRS, Marne-la-Vallée, France} 

\author{Anaël Lemaître}
\affiliation{Navier, Ecole des Ponts, Univ Gustave Eiffel, CNRS, Marne-la-Vallée, France} 

\author{Gianpietro Cagnoli}
\affiliation{Université de Lyon, Université Claude Bernard Lyon 1, CNRS, Institut Lumière Matière, F-69622 Villeurbanne, France} 

\author{Massimo Granata}

\author{Christophe Michel}
\affiliation{Laboratoire des Mat\'{e}riaux Avanc\'{e}s - IP2I, CNRS, Universit\'{e} de Lyon, Universit\'{e} Claude Bernard Lyon 1, F-69622 Villeurbanne, France} 

\author{Gianluca Gemme}
\affiliation{Istituto Nazionale di Fisica Nucleare, Sezione di Genova, via Dodecaneso 33, 16146 Genova, Italy} 

\author{Laurent Pinard}
\affiliation{Laboratoire des Mat\'{e}riaux Avanc\'{e}s - IP2I, CNRS, Universit\'{e} de Lyon, Universit\'{e} Claude Bernard Lyon 1, F-69622 Villeurbanne, France} 

\author{Maurizio Canepa}
\affiliation{OptMatLab, Dipartimento di Fisica, Universit\`{a} di Genova, via Dodecaneso 33, 16146 Genova, Italy} 

\begin{document}
	
	\newpage	
	
	\begin{abstract}
	Amorphous mixed titania-tantala coatings are key components of Bragg reflectors in the gravitational wave detectors (GWDs). Attaining the lowest possible values of optical absorption and mechanical losses in the coatings is of paramount importance for GWDs, and this requires a complex optimization of the coating deposition and post-deposition annealing. We present here a systematic investigation of the optical properties and internal friction of amorphous mixed titania-tantala coatings grown by ion beam sputtering. 
	We consider coatings with six different cation mixing ratios - defined as Ti/(Ti+Ta) - and we study them both in the as-deposited and annealed state. 
	All coatings have been subject to the same annealing of 500 $^\circ$C for 10 hours in air, which is the post-deposition treatment adopted so far for Bragg reflectors in GWDs applications.  
	By exploiting spectroscopic ellipsometry data and modelling, along with ancillary techniques, we retrieved the dielectric function of the coatings in a wide spectral range. When varying the mixing ratio and performing the annealing we find monotonic - in some cases, almost linear -  trends for most of the aforementioned properties. Remakably, the post-annealing Urbach energy displays a definite minimum for a mixing ratio around 20$\%$, very close to the composition of the coatings showing the lowest optical absorption for GWDs applications. We suggest that the observed minimum in the Urbach energy depends not only on the mixing ratio, but also on the annealing parameters. On the other hand, the minimum coating loss angle was found to be weakly dependent on the considered measurement frequency and to lie within a rather broad range of Ti content (cation ratios of 21\% and 44\%), suggesting that the search for an absolute minimum following post-deposition annealing should be rather sought in the study of the best annealing parameters for each specific cation ratio considered. This work constitutes a reference for the optical properties of the amorphous mixed titania-tantala coatings, and highlights the relevance of the Urbach energy in the optimization process of materials for high-performing Bragg reflectors.

		\vspace{1 cm}
		\emph{keywords: amorphous semiconductors, thin films, Urbach energy, internal friction, loss angle, ellipsometry, gravitational waves.}
		
	\end{abstract}

	\maketitle
	
	\section{Introduction}

	Amorphous oxides are key constituents of high-performance Bragg reflectors in technologies as diverse as optical atomic clocks\cite{RevModPhys.87.637}, passive optical gyroscopes\cite{Liu:19}, ultrastable frequency laser cavities\cite{PhysRevLett.118.263202} and Fabry-Perot cavities of present and future ground-based gravitational-wave detectors\cite{Acernese_2014,Aasi_2015,PhysRevD.88.043007,Maggiore_2020} (GWDs). In such advanced applications, optical multilayer coatings must display exceedingly low levels of optical absorption and thermal noise\cite{Granata_20_review}, as these two properties often set the limits of technological possibilities\cite{Bersanetti_2021}. While the precise microstructural features responsible for either optical absorption~\cite{Adachi_1999} or thermal noise~\cite{PhysRevB.93.014105} are still subject to discussion, it is widely expected that they originate from different types of defects of the amorphous structure. The best amorphous materials for high-performance Bragg reflectors, hence, should be in the most relaxed and lowest potential energy state.

    Optical multilayer coatings for GWDs are composed of alternating high- and low-refractive index amorphous oxides deposited at room temperature using ion beam sputtering (IBS)~\cite{Pinard_17}. Deposition is routinely followed by thermal annealing to decrease the internal stress and the optical absorption of the coatings~\cite{Granata_PRD_2017}. Crucially, annealing must avoid crystallization in order to preserve material homogeneity and isotropy, and to prevent optical scattering. The limits of the annealing temperature and duration are thus set by the material in the stack which crystallizes first~\cite{Granata_2020_mechanical}.
    
    Silica is the unchallenged low-index material for room-temperature applications as IBS deposition and annealing at moderate temperatures suffice to confer it excellent properties\cite{PhysRevMaterials.2.053607}. On the other hand, there is no evident solution of comparable quality for high-index layers. Good high-index materials, such as \ce{Ta2O5}, \ce{TiO2} or \ce{HfO2} present too high a thermal noise, as deposited, and meanwhile crystallize easily upon annealing: for these reasons, they cannot be used in pure form.  

    A strategy of primary importance towards the improvement of mirror coatings therefore consists in identifying adequate high-index \emph{mixed oxides}\cite{OIC_2003, Fazio_2022}. Indeed, it is known in many different contexts, ranging from metallic glasses to suspensions, that mixing strongly improves the stability of glasses against crystallization~\cite{Park_2015, lim_remarkably_2015,Abernathy_21}. The goal is then to optimize a complex trade-off between  refractive index, stability against crystallization, optical absorption, and thermal noise. This must be accomplished in consideration of the entire stack. In fact, thermal noise depends on the amount of materials used for the coatings and improving the index contrast between layers reduces the number of doublets while retaining the same reflectivity, thus contributing to reduce the overall losses and thermal noise.
    
    The simplest mixed oxides are typically envisioned as a combination of two pure oxides, one which is stable at the relevant annealing temperatures, and another one which has a higher index. 
    Important examples include \ce{TiO2}-\ce{Ta2O5}\cite{Harry_2006,Lee_06_1,Granata_2020_mechanical,Fazio_20}, \ce{SiO2}-\ce{TiO2}\cite{Chao_01}, \ce{TiO2}-\ce{Nb2O5}\cite{Amato_2021}, \ce{GeO2}-\ce{TiO2}\cite{Vajente21}, and \ce{ZrO2}-\ce{Ta2O5}\cite{Abernathy_21}.
    
    Today, the physical mechanisms at work in mixing remain to be clearly understood, with the consequence that the identification of optimal binary mixtures largely rests on educated guesses and a trial-and-error process~\cite{Fazio_20}. In particular, the oxide mixture currently used in GWDs mirrors is \ce{TiO2}-\ce{Ta2O5} with a Ti/(Ti+Ta) ratio around 20$\%$ because this was found to yield the lowest mechanical and optical losses after a thermal annealing of 500 $^\circ$C in air~\cite{Granata_2020_mechanical,Harry_2006}. However, the relation between the mixing ratios, the annealing parameters, and the resulting optical/mechanical properties of these mixtures, remain insufficiently characterized. 
    
    In this work, we study \ce{TiO2}-\ce{Ta2O5} coatings by systematically varying the cation mixing ratio (defined in percentage as Ti/(Ti+Ta)), before and after the standard annealing procedure for GWDs applications (10 hours at 500 $^\circ$C in air). We measure the dielectric function in a wide spectral range, accurately characterizing the fundamental absorption edge of the coatings. This edge is characterized not only by the energy gap, but also by a measure of the tail states within the gap - that is, the Urbach energy. This parameter is of particular relevance because it has been associated with topological or compositional disorder\cite{Zanatta_1996}, which must be kept at a minimum in the materials composing high-performance Bragg reflectors. All the optical properties considered in this work scale monotonously - with some tendence to linearity - with the cation ratio, save for the Urbach energy of the annealed coatings, which shows a very clear minimum for cation ratio around 20$\%$ - that is, the composition at which the lowest optical absorption has been measured in this kind of coatings\cite{Granata_20_review}. On the other hand, when measuring the mechanical loss, we find that the post-annealing values do not show a clear minimum, but are relatively low over a broad range of Ti/(Ta+Ti) ratios.

	\section{Materials and Methods}
	
	\subsection{Coating samples}
	Our samples consist in six sets of amorphous oxide coatings of varying cation ratio Ti/(Ta+Ti)$=0\%$, 21$\%$, 44$\%$, 71$\%$, 85$\%$ and 96$\%$, as measured through Rutherford backscattering spectrometry. They were deposited by IBS at the Laboratoire des Mat\'eriaux Avanc\'es\footnote{http://lma.in2p3.fr/} using a custom made coater machine, the so-called Grand Coater~\cite{Collaboration_2004}, developed to deposit the large high-reflective mirrors for GWDs.
	
	For optical investigation, in order to avoid backside reflections, the coatings were deposited on single-side polished silicon wafers. For mechanical measurements, which require low-loss substrates, they where deposited on silica disk-shaped resonators (2" or 3" of diameter and 1 mm thick). The Grand Coater enabled us to deposit these coatings at the same time on all the required substrates, thus guaranteeing that the optical and mechanical investigation are carried out on identical samples.
	
    All samples have been analysed as-deposited and after annealing at 500~\textcelsius~for 10 hours (ramp rate of 100~\textcelsius/hour), which is the post-fabrication annealing procedure for GWDs coatings so far adopted.
    
    Annealing induced the partial crystallization of the two coatings with the highest cation ratio (85$\%$ and 96$\%$). In this work we focus on the properties of amorphous coatings, therefore we analyzed these two samples only in the not-annealed state.

	\subsection{Optical Characterization}
	The optical characterization relied on spectroscopic ellipsometry (SE), which is especially well suited to study thin films~\cite{fujiwara2007spectroscopic,TompkinsHilfiker, Magnozzi_2018}, and which gives access to both electronic and optical properties via a model-based data analysis.
    
     Ellipsometry measurements were performed with a J.A. Woollam Co. Variable Angle Spectroscopic Ellipsometer. The considered spectral range was 190-2500 nm (0.5-6.5 eV), thus encompassing a large portion of the spectrum both below and above the energy gap. For each dataset the ellipsometry angles $\Psi$ and $\Delta$ plus the percentage of depolarization were acquired at three angles of incidence (AOI) (55$^\circ$, 60$^\circ$, 65$^\circ$); the spectral density was set to one datapoint each nm, totalling 2310 datapoints per spectrum. SE data were analyzed with the software WVASE 3.718 by J.A. Wollam Co.

     In order to model the SE data, we adapted a well-proven strategy based on the Cody Lorentz (C-L) model\cite{Ferlauto_2002}, which was designed to describe the optical response of amorphous semiconductors in a wide spectral range. In the C-L model, the imaginary part of the dielectric function reads:\cite{Amato_2019_optical}
	\begin{equation}
    \varepsilon_2=\begin{cases}
    L(E)G(E),&E> E_t\medspace,\\
    U(E),&0<E\leq E_t\medspace,
    \end{cases}
    \label{eq:Cody1}
    \end{equation}
    where $L(E)$ is the Lorentz oscillator and $G(E)$ describes the Cody behaviour for a parabolic electronic density of states
    \begin{equation}
    G(E)=\frac{(E-E_g)^2}{(E-E_g)^2+E_p^2}\medspace,
    \label{eq:Cody}
    \end{equation}
    where $E_g$ is the energy gap and $E_p$ is the limit energy separating the Cody behaviour from the Lorentz oscillator.
    $U(E)$ is related to the Urbach tails via the following equation,
    \begin{equation}
    U(E)=\frac{E_t L(E_t)G(E_t)}{E}e^\frac{E-E_t}{E_U}\medspace,
    \end{equation}
    where $E_t$ is the limit between the localized transitions of exponential tails and transitions involving only extended states and $E_U$ is the Urbach energy.
    In addition to the C-L model we also fitted the experimental SE data point by point, that is, without a parametric dispersion model. In a point-by-point fit, the calculated optical constants at different wavelengths are independent of one another\cite{SchubertIR}; this guarantees the best agreement with the experimental data, however the resulting optical constants are more noisy with respect to the ones obtained with a parametric model. We use the point-by-point fit as a further validation of the C-L model and results.\cite{Amato2020_correlation}.
    
	\subsection{Mechanical Loss Characterization}
	In amorphous materials, thermally driven random transitions cause a local variation of the elastic moduli of the film giving rise to dissipation. This mechanical dissipation is measured through the resonant method\cite{nowick1972anelastic} using a Gentle Nodal Suspension (GeNS)\cite{Cesarini_2009,GranataInternalFriction,Granata_PRD_2017}. In particular, at each cycle of vibration the dissipated energy $E_{\text{diss}}$ is related to the total energy stored in system $E_{\text{st}}$ through the relation
	\begin{equation}
	    E_{\text{diss}} = 2\pi\phi E_{\text{st}}\medspace,
	\end{equation}
	where $\phi$ is the so-called loss angle that quantifies the internal friction.
	The loss angle is obtained from the characteristic decay time $\tau_k$ of several resonance modes $k$ with frequency $\nu_k$ in the range $0.1-40$ kHz through the relation $\phi_k=(\pi\nu_k\tau_k)^{-1}$. 
	In composite resonators, made of substrate and coating, the coating loss angle is worked out by subtraction of the measurements of the bare substrate from that of the coated sample\cite{CAGNOLI20182165,Granata_2020_mechanical}.
	Once the coating loss was obtained, we used a power law to describe the frequency dependence across the measured frequency range. The power-law dependence of the loss angle is derived from an exponential behaviour of the barrier height in a two-level system (TLS)\cite{doi:10.1080/01418638108222343} as it has been confirmed in silica samples and over a wide temperature range\cite{Travasso_2007} and in Ti-doped tantala molecular dynamics simulations\cite{PhysRevB.93.014105}.
	
	In our study, for each Ti concentration, two nominally identical substrates have been coated at the same time in order to increase the statistics of the results and have a double check of the measured loss angles.
	
\section{Results}
\begin{figure}
		\centering
		\includegraphics[width=0.45\textwidth]{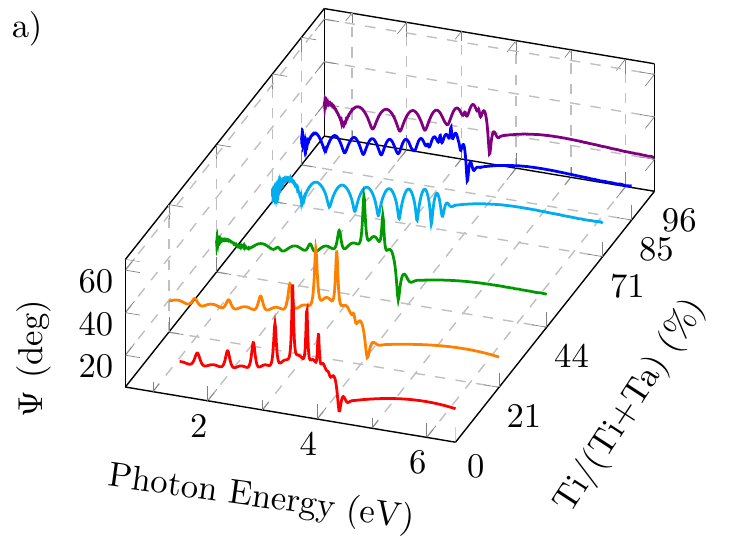}
		\includegraphics[width=0.45\textwidth]{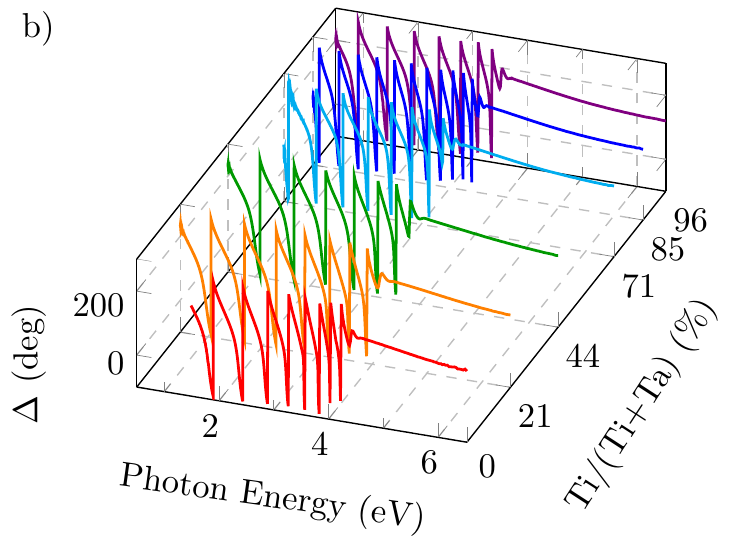}\\
		\includegraphics[width=0.92\textwidth]{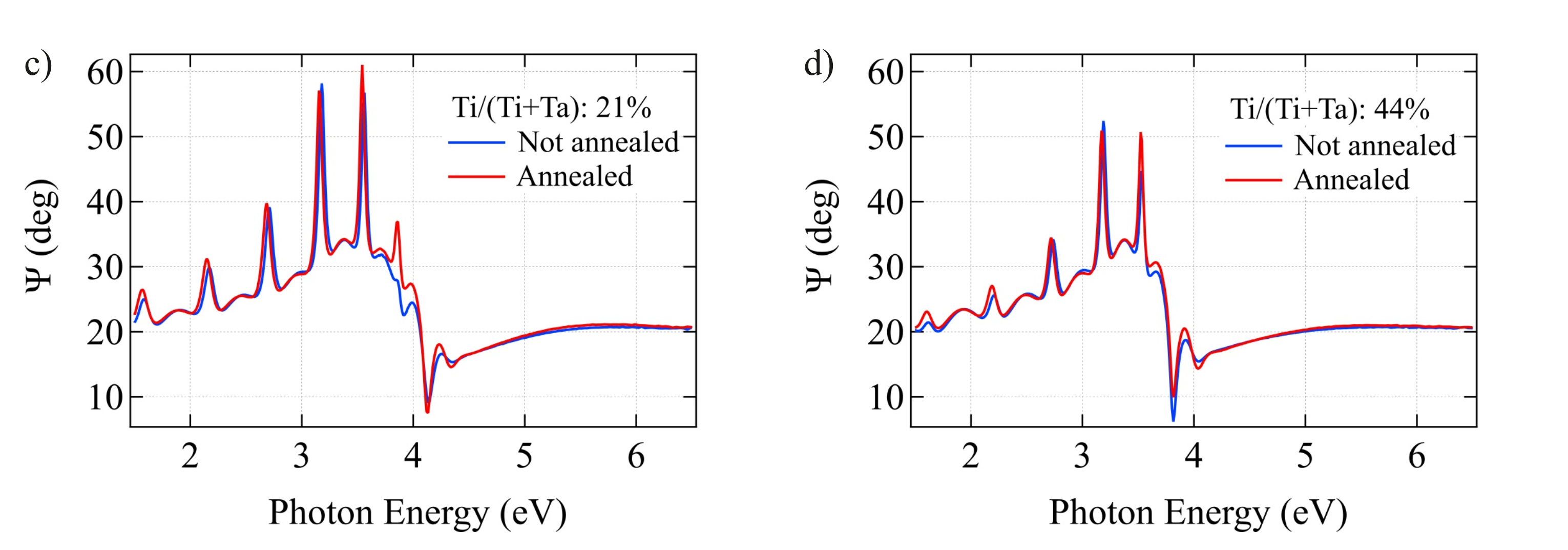}
		\caption{a), b): $\Psi$, $\Delta$ spectra measured on the as-deposited (\emph{i.e.}, not annealed) samples. The spectra are stacked according to the mixing ratio in each sample, expressed as percent Ti/(Ti+Ta).  c), d): $\Psi$ spectra of not annealed (blue curves) and annealed (red curves) coatings with cation ratio of 21$\%$ and 44$\%$, respectively. }
		\label{Fig1}
	\end{figure}

Figures~\ref{Fig1}a) and \ref{Fig1}b) show the ellipsometry data ($\Psi$ and $\Delta$) measured on the as-deposited (\emph{i.e.}, not annealed) samples. For the sake of clarity, only the data at 60$^\circ$ AOI are reported here. Generally speaking, each spectrum in Fig.~\ref{Fig1} exhibits three regimes depending on the spectral region. In the region of transparency (low photon energies) the spectra are characterized by periodic structures such as peaks or troughs due to the interference of multiple reflections inside the coating. In the region of strong absorption (high photon energies) the spectra are featureless. Between the two mentioned regions lies the onset of absorption. The 3D plots in Fig.~\ref{Fig1}a) and ~\ref{Fig1}b) allow to compare the features of the samples as the mixing ratio is varied. The most notable trend, which is visible both in the $\Psi$ and $\Delta$ spectra, is the redshift of the onset of absorption as the mixing ratio is increased. 
Figures ~\ref{Fig1}c) and ~\ref{Fig1}d) report the $\Psi$ spectra of the sample with Ti/(Ti+Ta) corresponding to 21$\%$ and 44$\%$, respectively, in the pristine (blue curves) and annealed (red curves) state. In both cases, the variations induced by annealing are mainly located in the spectral region corresponding to the onset of absorption; the largest of these variations (up to  9$^\circ$ in $\Psi$) are observed in the sample with cation ratio of 21$\%$.

\begin{figure}
\centering
\includegraphics[width=1\textwidth]{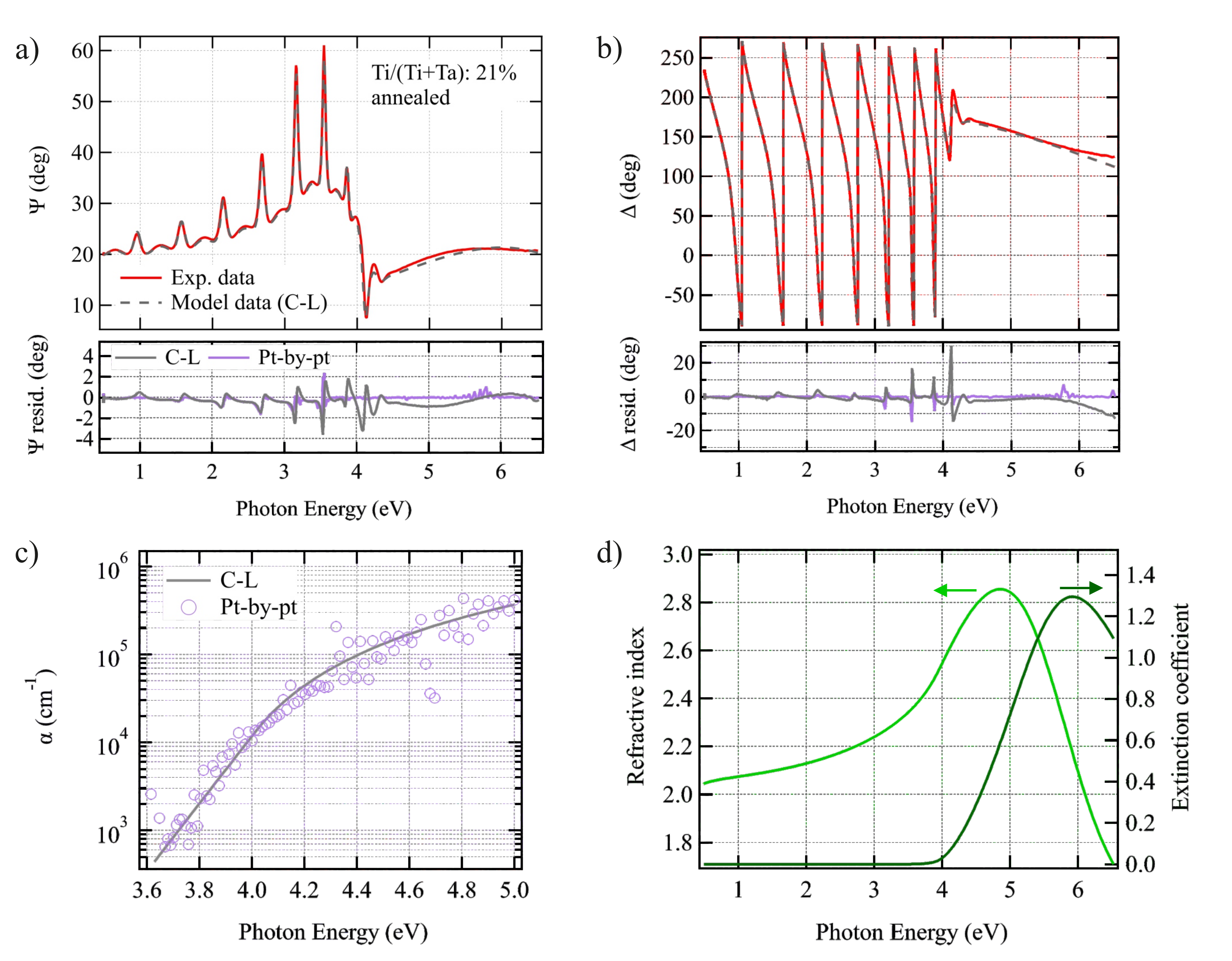}
\caption{Analysis of the SE data measured on the sample with Ti/(Ti+Ta): 21$\%$, annealed. a), b): comparison between experimental data and the results of Cody-Lorentz (C-L) model fit. Below the graphs are reported the residual from the C-L fit (gray lines) and from the point-by-point fit (violet lines). c): $\alpha$ obtained by C-L model (gray line) and point-by-point fit (violet circles). d): Refractive index (light green line) and extinction coefficient (dark green line) obtained from the C-L model.}
\label{Fig2}
\end{figure}

As an example of SE data modelling and fitting, we report in Fig.~\ref{Fig2} representative data from the sample with cation ratio of 21$\%$, annealed. Panels \ref{Fig2}a) and \ref{Fig2}b) report the $\Psi$ and $\Delta$ at 60$^\circ$ AOI, both experimental and calculated with the C-L model. Below those graphs are reported the fit residual obtained with the C-L fit and with the point-by point fit. The good agreement between these two data analysis methods can be appreciated in \ref{Fig2}c), where the absorption coefficient $\alpha$ is reported on a logarithmic scale. The refractive index and extinction coefficient of the coating, obtained with the C-L model, are reported in \ref{Fig2}d). 

\begin{figure}
	\centering
	\includegraphics[width=1\textwidth]{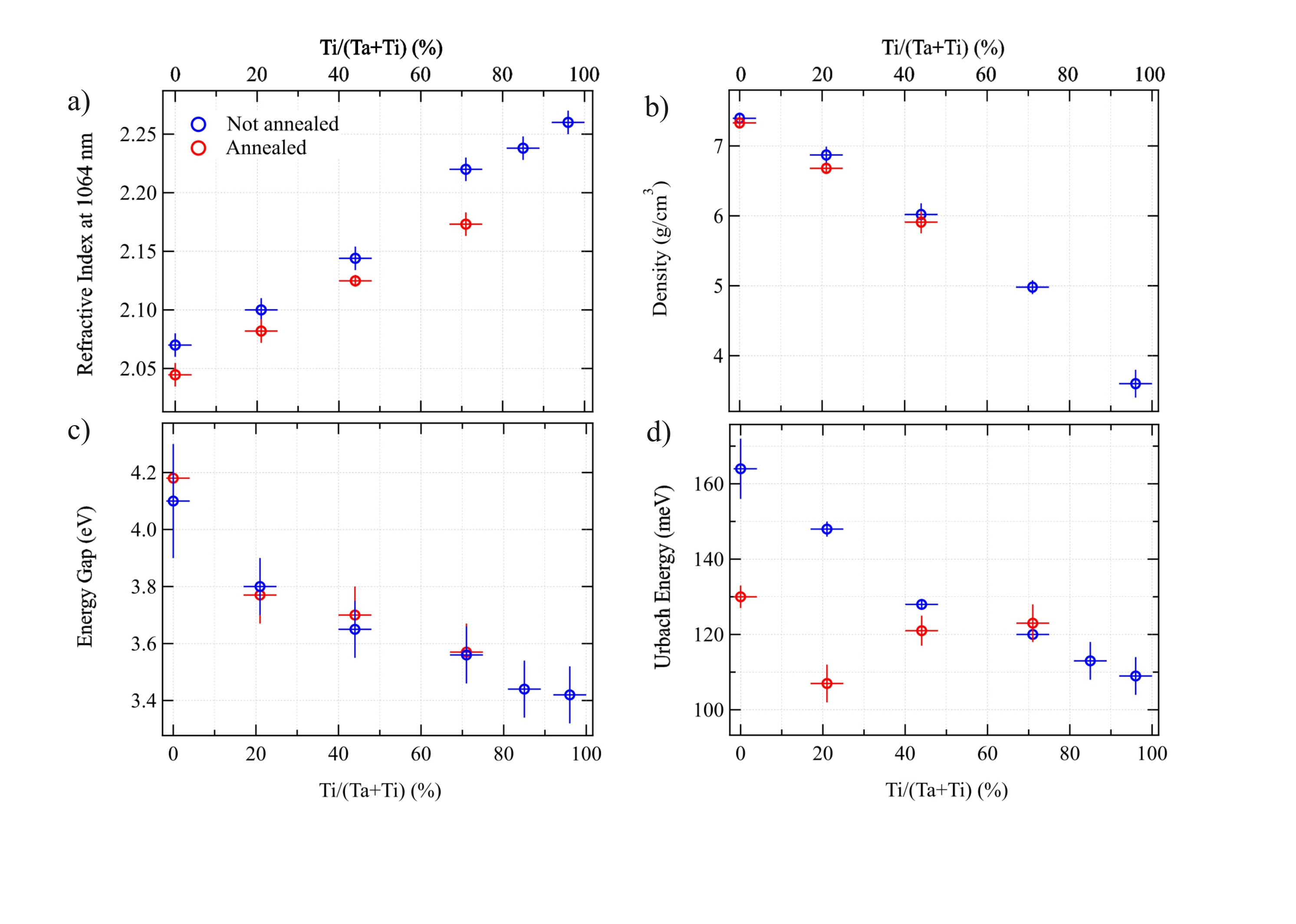}
	\caption{Properties of coatings as a function of the mixing ratio, determined from the analysis of ellipsometry data. Not annealed and annealed coatings are indicated with blue and red markers, respectively. a) Refractive index at $\lambda=1064$ nm. b) Density. The point marked with an asterisk is calculated exploiting the linear relation between refractive index and density in titania coatings, as reported in Ref. [\cite{Bundesmann2017}].  c) Energy gap. d) Urbach energy. Following the annealing protocol described in the text, the mixing ratio of 21$\%$ yields both the lowest Urbach energy and the largest annealing-induced decrease.}
	\label{Fig3}
\end{figure}

In Table~\ref{tab:thickness} and Fig.~\ref{Fig3} several properties of the films, resulting from the analysis of the SE data, are reported for varying the mixing ratio.
In the table, the thickness of coatings before and after the annealing are reported, along with the percent variation. The data show that the thickness of the films systematically increases as a result of the annealing; the percent variation decreases monotonically as the Ti content increases. 
In Fig.~\ref{Fig3}a) the refractive index at $\lambda=1064$ nm is seen to increase as the content of Ti atoms increases. Furthermore, the annealing reduces the refractive index by  $\sim$1-2\%, a finding compatible with the results previously reported on pure tantala coatings\cite{Anghinolfi_2013}.  
By combining the thickness data in Table~\ref{tab:thickness} with the measurements of the mass and diameter of the films, we calculated their density which is reported in Fig.~\ref{Fig3}b); the mass of the films remained constant upon annealing. The density of the coating decreases as the content of Ti atoms increases, as expected from the different atomic mass of the two transition metals. No direct measurement data were available on the samples with the highest Ti concentration, however an approximate value can be obtained by exploiting the linear relation between refractive index and density, which has been demonstrated for TiO$_2$ films\cite{Bundesmann_2017}. Based on such relation, having calculated $n$@550 nm = 2.39 for the not-annealed coating with 96$\%$ Ti, the density of that coating can be estimated as 3.6$\pm$0.2 g/cm$^3$ (point marked with an asterisk in Fig.~\ref{Fig3}b)). 

In Fig.~\ref{Fig3}c) the energy gap is reported, which shows a monotonic decrease as the Ti concentration increases and no clear variation trend induced by the annealing. Finally, in Fig. \ref{Fig2}d) the Urbach energy values are plotted. Interestingly, the Urbach energy trends for the pristine and annealed samples are qualitatively different. Pristine samples exhibit a monotonous decrease of the Urbach energy up to the highest values of Ti concentration; on the other hand, the datapoints for annealed samples indicate different Urbach energies, with a clear minimum at the mixing ratio of 21$\%$. At the same mixing ratio the largest difference of Urbach energy is observed between pristine and annealed samples. 

\begin{table}
\begin{tabular}{ >{\centering}m{2.8cm}  >{\centering}m{1.5cm} >{\centering}m{1.5cm} >{\centering}m{1.5cm} >{\centering}m{1.5cm} >{\centering}m{1.5cm} >{\centering\arraybackslash}m{1.5cm} }
\multicolumn{7}{c}{} \\
\hline
$\%$ Ti/(Ti+Ta) & 0 & 21 & 44 & 71 & 85 & 96\\
\hline
\hline
Thickness, not annealed (nm)&  579  & 504 & 480  & 580* & 817  & 532 \\
\hline
Thickness, annealed (nm)& 592 & 514 & 489  & 520*& 827** &  538** \\
\hline
$\%$ variation upon annealing & 2.2  & 2.0  & 1.9  & - & 1.2 & 1.1 \\
\hline
\end{tabular}
*different samples; **crystallized
\caption{\label{tab:thickness} Thickness of coatings determined by SE for pristine and annealed samples. The last row reports the percent variation in thickness as a consequence of the annealing.}
\end{table}

\begin{figure}
		\centering
		\includegraphics[width=0.45\textwidth]{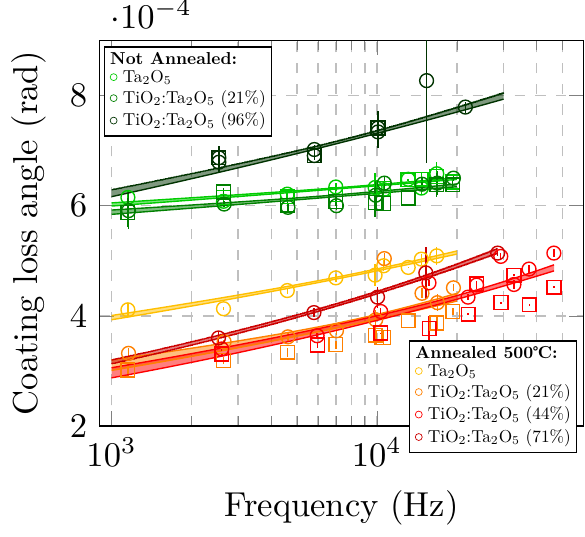}
		\caption{a) Coating loss angle as function of frequency of \ce{TiO2}:\ce{Ta2O5} with different cation mixing ratio, before and after the annealing. For each ratio, the squares and circles refer to measurements from two different disk-shaped resonators, coated at the same time and therefore nominally identical.}
		\label{Fig4}
	\end{figure}
	
In Fig.~\ref{Fig4} the coating loss angle data are plotted as a function of frequency. The loss angle is seen to increase with frequency for all samples. Although the slope varies, the loss angle is systematically smaller in the annealed samples; the lowest values are those corresponding to mixing ratios of 21$\%$ and 44$\%$, with no clear minimum associated to either of the two cases in the investigated frequency range.\\

\section{Discussion}

So far we have determined by measurements and data analysis how several properties in mixed titania-tantala coatings vary as a function of the mixing ratio and annealing. Here, we will attempt to link the observed variations and trends to their underlying physical causes, to draw comparisons with the existing reports in the literature where possible, and finally to extract useful information for optimizing the performance of this kind of coatings.

The interpretation of the dielectric function of mixed oxides is usually based on generalisations of the Clausius-Mossotti (CM) formula, resulting in the so-called "oxide additivity rule", which allows to predict the dielectric function (and therefore the square of the refractive index) of a very large number of binary and ternary metal oxide alloys, in general with an accuracy better than a few percent~\cite{doi:10.1063/1.1375806}. 
Therefore, the CM formula expresses the square of the refractive index of the mixed oxide  \ce{TiO2}:\ce{Ta2O5} as a linear combination of the polarizabilities of the single oxides \ce{Ta2O5} and \ce{TiO2}. Our data in Fig.~\ref{Fig3}a) show a monotonic relation between refractive index and mixing ratio, and are thus consistent with the aforementioned rule.  

The relation between density and mixing ratio (Fig.~\ref{Fig3}.b)) is clearly monotonic, which results from the atomic mass contrast between the two transition metals ($m_{a,(Ti)}=47.9$u and $m_{a,(Ta)}=180.9$u). The relation, however, is only approximately linear for non-annealed samples. Annealing induces a slight ($\sim$1-2\%), but measurable reduction of the film density. This effect is better seen from the associated changes in film thickness in Table~\ref{tab:thickness}, from which we observe that the relative change is smaller at higher Ti content. 
As for pure tantala coatings, our findings are fully compatible with previous reports both from the standpoint of the absolute density, 7.4$\pm$0.1 g/cm$^3$, which compares with Ref. [\citen{Alderman_2018}], and with regard to the density change upon annealing for films of similar thickness ($\sim$1-2\% in Refs. [\citen{Paolone_2021,Granata_2020_mechanical,Amato_2019_annealing,Anghinolfi_2013}]).

We shall now consider the energy gap, displayed in Fig.~\ref{Fig2}c). The overall trend for varying cation ratio is approximately linear, with the exception of the pure tantala data points; this observation is consistent with that reported by Fazio et al.\cite{Fazio_20}. Unlike refractive index and density, the energy gap is not measurably affected by annealing: in Fig.~\ref{Fig2}c), the annealed and non-annealed data points remain consistent within error bars, regardless of the mixing ratio. This, again, is consistent with, and extends, previous results on pure tantala coatings with similar thicknesses, the energy gap of which was found not to be affected by annealing up to 600 $^\circ$C\cite{Anghinolfi_2013}. Moreover, this insensitivity with annealing is in line with the idea that the gap is controlled by the nearest neighborhood of metallic atoms, but not by broader structural features (medium range order) that are the most affected by structural relaxation\cite{Thapa_2022}.

The data discussed so far show a monotonic dependence on the cation ratio, which in some cases is even compatible with a linear relation. For this reason, the trends of refractive index, density, and energy gap in this kind of amorphous mixed oxides coatings can generally be understood in terms of the aforementioned additivity rule. 
The Urbach energy, reported in Fig.~\ref{Fig2} d), shows instead a quite different picture. Its trend is still monotonic in the as-deposited coatings, but shows a clear minimum in the annealed ones, corresponding to the cation ratio of 21$\%$. Moreover, for that mixing value, the Urbach energy is decreased by $\sim$40 meV upon annealing, which is the largest reduction among the investigated samples. 
Therefore, the Urbach energy manifests two very different trends before and after the annealing. In the first case, it seems to be governed by the mixing ratio, in other words: the higher the Ti content, the lower the Urbach energy. In the second case, we register a significant decrease in the Urbach energy for mixing ratio 0$\%$ and 21$\%$, but very small and no variation at all for mixing ratio 44$\%$ and 71$\%$, respectively. This means that the  annealing parameters employed so far (10 hours at 500 \textcelsius) can have large effects on the Urbach energy, but only for certain mixing ratios. 

In order to explain for this peculiar behavior, it is useful to recall that from an atomistic point of view, Urbach tails have been related to \emph{extended} structural patterns, involving a correlation of interatomic bond lengths or angles overal many interatomic distances\cite{Pan_2008, Inam_2010, Drabold_2011}. According to this model, Urbach tails are mainly related not only to the short range but also to the intermediate range. Using a combination of experimental and modelling tools, Prasai et al. demonstrated that a thermal annealing in amorphous zirconia-tantala coatings leaves the short-range order almost unchanged, but has a significant effect on the intermediate-range order\cite{Prasai_2019}. This explains that annealing can have large effects on the Urbach energy of amorphous mixed oxide coatings as already shown elsewhere\cite{Amato_2019_annealing}. Crucially, however, our data show that these effects depend on the mixing ratio, as they are sizable for low Ti content, but very small or negligible for medium Ti content. 
We hypothesize that the observed behavior is determined by the fact that for medium Ti content, the maximum temperature (and/or time duration) of the annealing are not sufficient to remove the extended structural patterns mentioned at the beginning of this paragraph. Indeed, the same annealing parameters can have profoundly different effects on the microscopic structure of the coatings with different mixing ratios - in the extreme case, the same annealing can lead the coatings with high Ti content to crystallization, while leaving those with low Ti content perfectly amorphous. Following our hypothesis, we suggest that the minimum observed in the Urbach energy for cation ratio of 21$\%$ could not be the one and only for this kind of materials - other minima could be found at different cation ratios \emph{and} with a different annealing protocol (\emph{i.e.}, with a different combination of maximum temperature and time duration). 
Interestingly, the minimum of the Urbach energy reported in this work is mirrored by the minimum in the optical absorption measured in the region of transparency (1064 nm) in similar coatings\cite{Fazio_20}; this implies that both the extended disorder (which gives rise to the Urbach tails) and the localized, individual defects (which give rise to optical absorption deep within the energy gap) are minimized in annealed titania-tantala coatings with mixing ratio around 20$\%$.

The available data show that annealing induces a significant reduction of the loss angle, a finding which confirms the previous results from the literature\cite{Amato2020_correlation,Fazio_20}. However, unlike the Urbach energy, the measured loss angles in the annealed samples do not have a well-defined minimum. In fact, the loss angle data-points relative to the annealed 21$\%$ and 44$\%$ samples lie very close to each other, one being slightly lower than the other depending on the considered frequency. This further strengthens our hypothesis that the search for an absolute minimum following post-deposition heat treatment must be sought in the individual study of the best annealing for each different type of titania-tantala mixing concentration - a study in which temperature-depentent, \emph{in situ} SE is a valid research tool\cite{Magnozzi_2017, Zollner_2022}. Furthermore, considering the frequency trend of the loss angle, the search for a minimum after annealing may depend on the frequency range of interest. 

\section{Conclusions}
	
We have systematically studied the dielectric function and loss angles in amorphous titania-tantala coatings, determining how they vary as a function of the cation mixing ratio and after a thermal annealing.
We found that the aforementioned properties scale monotonously with the mixing ratio, with the notable exception of the Urbach energy of the annealed samples which has a clear minimum at the cation ratio of 21$\%$. A minimum in the Urbach energy can be correlated to the least amount of extended defects in the amorphous structure, which is a potentially good configuration for a coating material in GWDs applications. Indeed, a very similar mixing ratio was found to yield the lowest optical absorption in the transparency region. On the other hand, our data on mechanical losses show no clear minimum for intermediate Ti content, a feature that can be ascribed to the choice of the annealing parameters which might be different for obtaining the lowest loss at each Ti content. Therefore, we suggest that the minimum we found in the Urbach energy is not the one and only possibile in amorphous titania-tantala coatings - other minima could be found at different mixing ratios, depending on the annealing protocol. 
This work constitutes a reference for the optical properties of amorphous titania-tantala coatings and proposes the Urbach energy a parameter to guide the optimization process of amorphous mixed oxide coatings for high-performance Bragg reflectors.


\section*{Conflicts of interest}
There are no conflicts of interest to declare.
	
\section*{Acknowledgments}
A subset of the optical data contained in this manuscript was included in the Technical Summary of the Optical Interference Coatings 2022 conference. This work was supported by the Virgo Coating R$\&$D (VCR$\&$D) Collaboration. We acknowledge the support of Italian Ministry for University and Research (MUR) through the project 'Dipartimenti di Eccellenza 2017-2022' (DIFILAB). 
	
\bibliography{Bibliography}

\end{document}